# A single-photon switch and transistor enabled by a solid-state quantum memory

Shuo Sun[1], Hyochul Kim[1], Zhouchen Luo[1], Glenn S. Solomon[2], and Edo Waks[1,*]

**Affiliations:**

[1]Department of Electrical and Computer Engineering, Institute for Research in Electronics and Applied Physics, and Joint Quantum Institute, University of Maryland, College Park, Maryland 20742, USA

[2]Joint Quantum Institute, National Institute of Standards and Technology, and University of Maryland, Gaithersburg, Maryland 20899, USA

*Correspondence to: edowaks@umd.edu

**Abstract:**

Single-photon switches and transistors generate strong photon-photon interactions that are essential for quantum circuits and networks. However, to deterministically control an optical signal with a single photon requires strong interactions with a quantum memory, which have been lacking in a solid-state platform. We realize a single-photon switch and transistor enabled by a solid-state quantum memory. Our device consists of a semiconductor spin qubit strongly coupled to a nanophotonic cavity. The spin qubit enables a single gate photon to switch a signal field containing up to an average of 27.7 photons, with a switching time of 63 ps. Our results show that semiconductor nanophotonic devices can produce strong and controlled photon-photon interactions that could enable high-bandwidth photonic quantum information processing.

**Main text:**

Photons are ideal carriers of quantum information, but the lack of deterministic photon-photon interactions have limited their applications in quantum computation and quantum networking. Recent advances in strong light-matter interactions using neutral trapped atoms (*1-5*) have enabled optical nonlinearities operating at the fundamental single-photon regime. But neutral atoms require large and complex laser traps and operate at low bandwidths on the order of Megahertz, making them challenging to integrate into compact devices. Circuit quantum electrodynamics systems also support strong nonlinearities (*6, 7*). But they operate only in the microwave regime and are difficult to scale to optical frequencies. The realization of a compact solid-state single-photon nonlinearity at optical frequencies remains a key missing ingredient for scalable chip-integrated quantum photonic circuits.

Nanophotonic structures coupled to quantum emitters offer an attractive approach to realize single-photon nonlinearities in a compact solid-state device. However, most of previous works utilized quantum emitters that act as two-level atomic systems (*8*), which is fundamentally limited by a time-bandwidth tradeoff that makes deterministic single-photon switching impossible (*9, 10*).

A quantum memory can overcome this limit, enabling a single photon to deterministically switch a second photon (*11*). It can also realize a single-photon transistor where one photon can switch a signal containing multiple photons (*12*), a crucial building block for scalable quantum circuits (*13*). Recently there has been great progress in controlling photons with solid-state qubits (*14-16*), as well as controlling a solid-state qubit with a photon (*17*). However, neither a single-photon switch nor a single photon transistor has been realized using a solid-state quantum memory.

In this letter, we report a single-photon switch and transistor enabled by a solid-state spin qubit coupled to a nano-cavity. Our spin qubit is composed of a single electron in a charged quantum dot. Figure 1A shows the quantum dot level structure, which includes two ground states with opposite electron spin that form a stable quantum memory, labelled as $|\uparrow\rangle$ and $|\downarrow\rangle$, and two excited states that contain a pair of electrons and a single hole with opposite spins, labelled as $|\uparrow\downarrow,\Uparrow\rangle$ and $|\uparrow\downarrow,\Downarrow\rangle$. Figure 1B shows a scanning electron microscope image of a fabricated cavity (*18*). We attain spin-dependent coupling by applying a magnetic field of 5.5 T along the growth plane of the device (Voigt configuration). At this magnetic field, transition $\sigma_1$ is resonant with the cavity mode while all other transitions are detuned. Using cross-polarized reflectivity measurements, we determine the coupling strength $g$, cavity energy decay rate $\kappa$, and transition dipole decoherence rate $\gamma$ to be $g/2\pi = 10.7 \pm 0.2\,\text{GHz}$, $\kappa/2\pi = 35.5 \pm 0.6\,\text{GHz}$ and $\gamma/2\pi = 3.5 \pm 0.3\,\text{GHz}$ respectively (*18*), which puts the device at the onset of the strong coupling regime defined by the condition $g > \kappa/4$ (*8*).

Figure 1C shows the working principle of the single-photon switch and transistor. A gate pulse first sets the internal quantum memory of the switch. If the gate pulse contains zero photons, the spin stays in the spin-down state. But if the gate pulse contains one photon, it sets the spin to spin-up. Subsequently, the spin-state controls the cavity reflection coefficient, thereby changing the polarization of reflected signal photons. To implement these two steps, we use the pulse sequence shown in the inset. We prepare the quantum dot in a superposition of its spin ground states given by $(|\uparrow\rangle+|\downarrow\rangle)/\sqrt{2}$ using an initialization pulse to optically pump the spin to spin-down, followed by an optical rotation pulse that creates a $\pi/2$ spin rotation (*19*). The system then freely evolves for a time $\tau$, followed by a second identical rotation pulse. We inject the gate pulse between these two spin rotation pulses. If we set the free evolution time to be an integer number plus one half of spin precession period, then in the absence of a gate photon the spin will evolve to the state $(|\uparrow\rangle-|\downarrow\rangle)/\sqrt{2}$ and the second rotation pulse will rotate it back to the spin-down state. But if a single gate photon reflects from the cavity, it applies a relative π phase shift between the spin-up and spin-down state, which reflects the spin along the *x*-axis of the Bloch sphere. In this case the second rotation pulse rotates the spin to the spin-up state. A signal field then reflects off the cavity and undergoes a spin-dependent polarization rotation. The supplementary text provides a detailed analytical derivation of each step of the device operation.

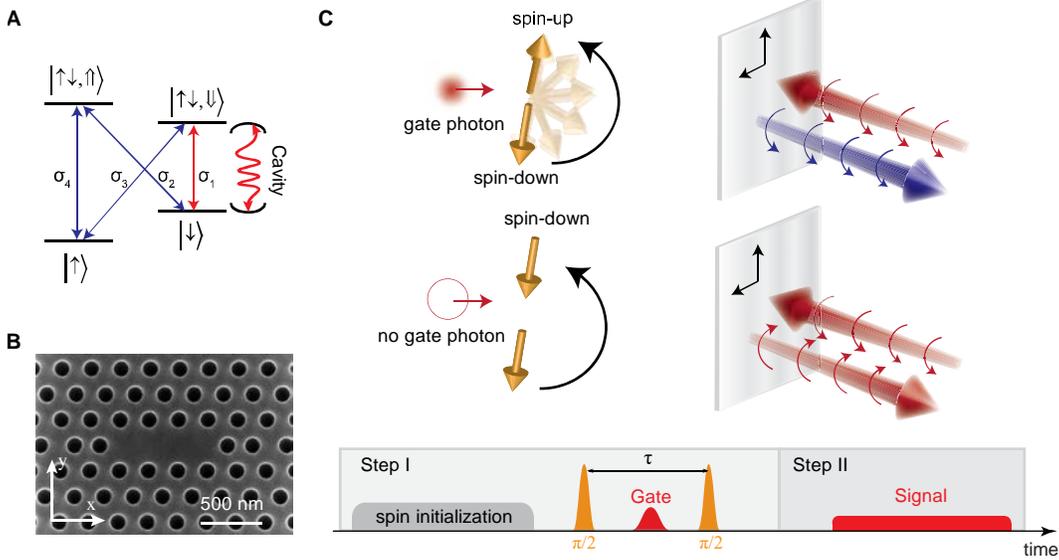

**Fig. 1: Schematics of a single-photon switch and transistor.** (**A**) Energy level structure of a charged quantum dot in the presence of a magnetic field applied in the Voigt geometry. (**B**) Scanning electron microscope image of a fabricated photonic crystal cavity device. (**C**) Schematic working principle of the single-photon switch and transistor, along with pulse timing diagram.

We prepare the pulse sequence in Fig. 1C using a pair of synchronized mode-locked lasers and an amplitude modulated external cavity laser diode (*18*). We first characterize the switching behavior of the device using a signal field that has an average photon number per pulse of $N_s = 0.42 \pm 0.05$ contained within the transverse spatial mode of the cavity (see supplementary text for measurement of signal photon number). We prepare the incident signal field in the right-circular polarization, and measure the intensity of the left-circular polarization component of the reflected signal field using a fixed polarizer after the cavity. Figure 2A shows the measured transmittance of the signal field passing through the polarizer as a function of $\tau$ in the absence of the gate pulse (see supplementary text for extraction of transmittance from measured intensities). We define the transmittance contrast as $\delta = T_{up} - T_{down}$, where $T_{up}$ and $T_{down}$ are the transmittance of the signal field when we prepare the spin to spin-up and spin-down respectively with the two rotation pulses, corresponding to the maximum and minimum transmittance in the oscillation. From the numerical fit (solid line), we calculate the transmittance contrast to be $\delta = 0.24 \pm 0.01$. This value differs from the ideal contrast of unity due to both imperfect spin fidelity of $F = 0.78 \pm 0.01$ after the two rotation pulses, and a finite cooperativity of $C = 2g^2/\kappa\gamma = 1.96 \pm 0.19$.

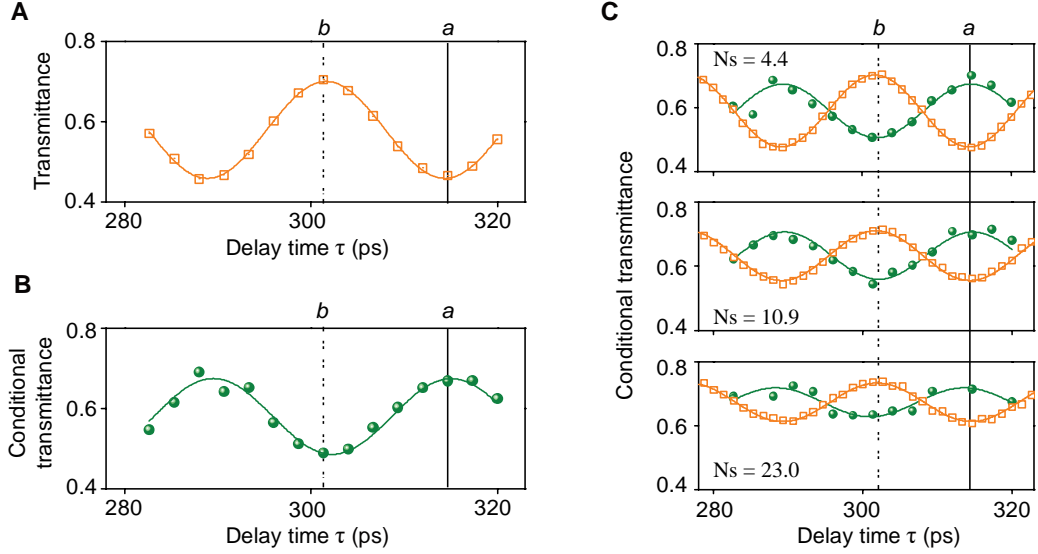

**Fig. 2: Demonstration of a single-photon switch and transistor.** **(A)** Transmittance of the signal field in the absence of the gate field as a function of delay time τ between two spin rotation pulses. **(B)** Transmittance of the signal field conditioned on a detecting a gate photon as a function of delay time τ between two spin rotation pulses. **(C)** Conditional transmittance of the signal field as a function of delay time τ between the two spin rotation pulses with (green) and without (orange) a gate photon when we set the average signal photon number per pulse to be $4.4 \pm 0.5$, $10.9 \pm 1.2$, and $23.0 \pm 2.5$ respectively.

Figure 2B shows the case where we inject a 63-ps gate pulse containing an average of 0.21 photons per pulse coupled to the cavity. To demonstrate that a single gate photon controls the transmittance of the signal field, we perform a two-photon coincidence measurement between the gate and signal photons (*18*). The green circles show the measured signal transmittance conditioned on detecting a reflected gate photon as a function of $\tau$, and the green solid line shows a numerical fit to the same model used in Fig. 2A. The oscillations shift by π due to spin-flips induced by a single gate photon.

The vertical solid line (labeled as "*a*") in Fig. 2A and 2B indicates the condition where the spin undergoes an integer plus one half number of rotations around the Bloch sphere during its free evolution time. At this condition a gate photon causes the polarization of the signal field to rotate and preferentially transmit through the polarizer as described in Fig. 1C. The vertical dashed line "*b*" shows a second operating condition that also leads to optimal switching operation. This condition corresponds to the reverse switching behavior where the gate photon prevents the signal field polarization from rotating, thereby decreasing the transmittance. At both conditions, the gate pulse induces a change in the signal transmittance by $0.21 \pm 0.02$. For an ideal gate pulse containing a single photon, the transmittance change should be equal to the transmittance contrast of 0.24 calculated in Fig. 2A. In our case the change in transmittance is slightly degraded because we use an attenuated laser to produce the gate pulse, which has a small probability of containing multiple photons. We define the switching contrast $\xi$ as the change in the transmittance induced by a single gate photon. By correcting for multi-photon events in the gate field (see supplementary text), we attain $\xi = 0.24 \pm 0.02$, which matches the transmittance contrast $\delta$.

The quantum memory in our device has a lifetime that is significantly longer than the bandwidth of the switch. Thus, once a gate photon sets the memory state, the device can switch many signal photons before the spin state decays. This property enables a single photon transistor where a single gate photon can control a signal composed of many photons, a significant distinction from switches lacking a quantum memory (9, 10). Figure 2C shows the transmittance of the signal field as a function of delay time $\tau$, where the average number of signal photons $N_s$ per pulse is set to be $4.4 \pm 0.5$, $10.9 \pm 1.2$, and $23.0 \pm 2.5$ respectively. The green circles show the transmittance conditioned on detecting a gate photon, and the orange squares show the transmittance without the gate pulse. The green and orange sold lines show the numerical fits to the same theoretical model used in Fig. 2A and 2B. The transmittance shows clear switching behavior for all cases. We calculate the switching contrast at the three signal photon numbers to be $\xi = 0.22 \pm 0.03$, $\xi = 0.17 \pm 0.02$ and $\xi = 0.12 \pm 0.02$ respectively.

The switching contrast degrades with increasing signal photon number because each signal photon can apply a back action on the spin through inelastic Raman scattering with a small probability, inducing an undesired spin-flip that resets the state of the internal quantum memory. This weak back-action limits the number of signal photons that can reflect from the cavity before a spin-flip event resets the spin-state. The blue circles in Fig. 3A show the measured transmittance contrast in the absence of the gate pulse as a function of the average number of photons in the signal field. This contrast quantifies the degree of self-switching induced by the signal without a gate field. The solid line shows a numerical fit of the data to an exponential function of the form $\exp(-N_s/N_{avg})$, where $N_{avg}$ is the average number of signal photons it takes to flip the spin. From the fit we determine $N_{avg} = 27.7 \pm 8.3$.

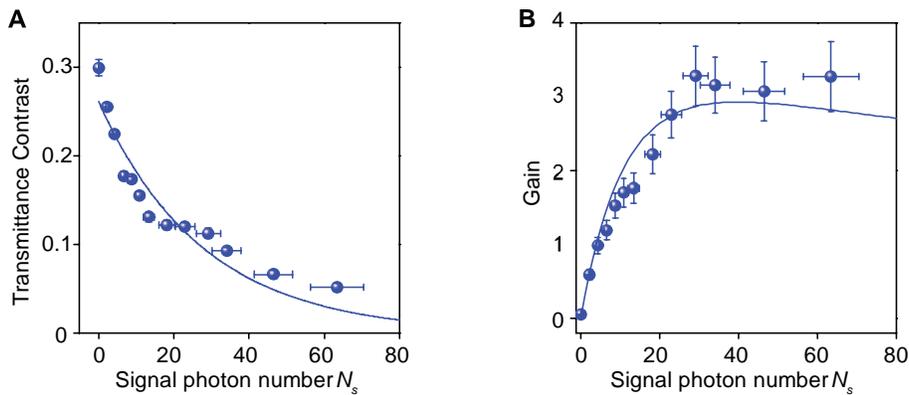

**Fig. 3: Gain of the single-photon transistor.** (**A**) Transmittance contrast as a function of average signal photon number. (**B**) Transistor gain as a function of average signal photon number.

An important feature of transistors is that they exhibit a gain above unity. We define the device gain as the change in the average number of transmitted signal photons induced by a single gate photon (*2, 3*). We determine the gain using the difference in the number of transmitted signal photons when the spin is coherently prepared in spin-up and spin-down states respectively, which is given by $G = N_s \delta$. The blue circles in Fig. 3B show the gain as a function of average signal photon number. The gain of the transistor increases initially, but saturates at strong signal fields due to an increased probability of spin-flip from inelastic scattering. We achieve the maximum gain of $G = 3.3 \pm 0.4$ with a photon number of $N_s = 29.2 \pm 3.2$. The blue solid line shows the numerically calculated gain, which agrees well with measurements.

In conclusion, we demonstrated a single-photon switch and transistor enabled by a solid-state quantum memory. The strong light-matter coupling strength in the nanophotonic device enables switching on picosecond timescales. In our current device, we excite and collect from the out-of-plane direction, which results in low coupling and collection efficiency that limits the usable gain. A scalable device suitable for quantum information processing will require much higher efficiencies, as photon loss constitutes a dominant error mechanism for photonic qubits. Recently there has been significant progress in improving coupling efficiency of nanophotonic devices, including using adiabatic tapered structures to directly couple to fiber (*20*), adopting designs that have better spatial mode-matching with a fiber (*21, 22*), or by coupling directly to on-chip waveguides (*23*). On-chip tuning (*24*) or hybrid integration techniques (*25, 26*) could further enable integration of multiple qubits and cascaded devices. Ultimately, such a device could enable a variety of important applications using compact chip-integrated platforms, including low energy electro-optics (*27*), photonic quantum circuits (*28*), non-destructive photon detection (*29*), and scalable quantum repeaters (*30*) for quantum networks (*31*).

**Acknowledgments:** The authors would like to acknowledge Gerry Baumgartner and Mark Morris at the Laboratory for Telecommunication Sciences for providing us with superconducting nanowire single-photon detectors. This work was supported by the Physics Frontier Center at the Joint Quantum Institute, the National Science Foundation (grant number PHY1415485 and ECCS1508897), and the ARL Center for Distributed Quantum Information.

**Author contributions:** S.S. and E.W. conceived and designed the experiment, prepared the manuscript, and carried out the theoretical analysis. S.S. carried out the measurement and analyzed the data. H.K. contributed to sample fabrication. Z.L. contributed to optical measurement. G.S.S. provided samples grown by molecular beam epitaxy.

**Competing interests:** The authors declare no competing financial interests.

**Materials and Methods**

1. Device design and fabrication

We start device fabrication with an initial wafer composed of a 160-nm-thick GaAs membrane grown on top of a 900-nm-thick $Al_{0.78}Ga_{0.22}As$ sacrificial layer. The GaAs membrane contains a single layer of InAs quantum dots at its center (density of 10 - 50/$\mu m^2$). Due to residual doping background, a fraction of quantum dots naturally confines an additional electron, which can be further stabilized by a weak He-Ne laser illumination. We fabricate photonic crystal structures using electron-beam lithography, followed by inductively coupled plasma dry etching and selective wet etching of the sacrificial AlGaAs layer. The cavity design is based on a three-hole defect in a triangular photonic crystal (*32*), with a lattice constant of 240 nm and a hole radius of 72 nm. The cavity is single sided due to a distributed Bragg reflector composed of 10 layers of GaAs and AlAs grown below the sacrificial layer.

2. Experimental Setup

2.1 Complete Schematic of the experimental setup

Figure S1 shows the schematic illustration of the whole experimental setup. We mount the sample in a closed-cycle cryostat and cool it down to 3.6 K. We use a superconducting magnet to apply magnetic fields up to 9.2 T along the in-plane direction (Voigt configuration). We use a confocal microscope with an objective lens that has a numerical aperture of 0.68 to perform sample excitation and collection. We prepare the polarization of all the incident light to right-circular polarization using a quarter-wave plate and a polarizer, including optical spin initialization, rotation, and the gate and signal pulses. We split the reflected signal into two different paths to collect the gate and signal field respectively. We set the detection polarization basis for the signal path to be left-circular polarization (cross-polarization) by using another set of quarter-wave plate and polarizer. We finely tune the detection polarization for the gate path to achieve the optimal switching contrast by using a set of quarter-wave plate, half-wave plate, and polarizer. Note that instead of preparing the incident gate pulse in the correct polarization basis that coherently flips the spin (which should be along the cavity polarization), we measure the gate pulse in this polarization basis after reflection to post-select the gate photons that couple to the cavity. We focus the collected photons into single-mode fibers that spatially filter out spurious surface reflection. The collected signal is sent to either a grating spectrometer with a resolution of 7 GHz for spectral measurement, or two superconducting nanowire single-photon detectors with a time resolution of 100 ps for photon correlation measurements. When sent to single-photon detectors, we utilize etalon filters with 0.3 nm bandwidth and fiber Fabry-Perot filters with 8 GHz bandwidth to spectrally reject the reflected optical pumping and rotation lasers from the sample surface.

2.2 <u>Pulse generation and synchronization</u>

We use two mode-locked Ti: Sapphire lasers (referred as master and slave) and a continuous-wave external cavity diode laser (linewidth below 300 kHz) followed by a Lithium Niobate electro-optic modulator (bandwidth >10 GHz, extinction ratio >40 dB) to generate all the pulses. We synchronize the slave laser to the clock of the master laser by piezo feedback in its cavity. A phase-lock loop in the synchronization electronics allows fine tuning of their relative delay. To synchronize the pulses generated from the electro-optic modulator, we use the clock of the master laser to trigger the signal delay generator that drives the electro-optic modulator. The signal delay generator produces a single TTL pulse with an electrically controllable delay with respect to each trigger pulse.

Figure S2 shows the setup to generate the pulse sequence that we used for the measurement in Fig. 2A and 2B of the main text. We use the 2-ps pulse from the master laser followed by a spectrometer grating to generate a 63-ps-long pulse. We then split the pulse into two identical ones with a fixed delay of 1.5 ns using an optical interferometer, which serve as the gate and signal pulses respectively. Both the gate and signal pulses are resonant with transition $\sigma_1$ and the cavity. We use the 4-ps pulse from the slave laser for optical spin rotations. The rotation pulse is red detuned by ~500 GHz from the cavity resonance. We also split this pulse into two to implement the Ramsey interferometry. We mount one of the retroreflectors of the Ramsey interferometer on a movable translation stage to control the delay time $\tau$ between the two rotation pulses. We use the electro-optic modulator to generate a 4-ns optical pumping pulse. The optical pumping pulse is resonant with transition $\sigma_4$ of the quantum dot.

Figure S3 shows the setup to generate the pulse sequence that we used for the measurement in Fig. 2C of the main text. The setup is almost identical with Fig. S2, except that we now utilize the electro-optic modulator and the continuous wave laser to generate a longer signal pulse. The inset shows the measured signal pulse shape, which does not follow a square shape since we are operating at the bandwidth limit of the signal delay generator. The generated pulse could be well characterized by a Gaussian function with a full width at half maximum (FWHM) of 1.34 ns (see inset at the bottom right corner). Such a longer signal pulse allows us to inject more photons without violating the weak excitation assumption. Since we only have one electro-optic modulator, we now use the 2-ps master laser to generate both the gate pulse and the optical pumping pulse. We split the 2-ps master laser into two paths, one going through a spectrometer grating to generate the 63-ps gate pulse, and the other one going through a 0.9 GHz fiber Fabry-Perot tunable filter to generate a 500-ps optical pumping pulse. Note that the center frequencies of the two spectral filters are resonant with transition $\sigma_1$ and $\sigma_4$ respectively. We do not observe any degradation of switching contrast when we switch to this shorter optical pumping pulse.

2.3 Two-photon coincidence measurements

To obtain the green circles shown in Fig. 2B and 2C, we perform two-photon coincidence measurements at each delay time $\tau$. This measurement records the number of transmitted signal photons conditioned on detection of a gate photon. Figure S1 shows the measurement setup. We split the reflected light into two paths with different detection polarization bases, designed to collect the gate and signal fields respectively. We use two superconducting nanowire single-photon detectors to collect photons from each path, and correlate their detections using a time correlated single-photon counting module. The blue line in Fig. S4 shows one of the fifteen coincidence histograms we obtained to extract the data shown in Fig. 2B of the main text. We observe three distinct peaks. The central peak at zero delay-time corresponds to the coincidence between the detection of gate photons at both the gate and signal paths, or the detection of signal photons at both the gate and signal paths. The left peak at minus delay-time corresponds to the case where the gate path detects a signal photon, and the signal path detects a gate photon. The right peak at plus delay-time corresponds to the case where the gate path detects a gate photon, and the signal path detects a signal photon. The separation between each peak is 1.5 ns, corresponding to the delay time between the gate and signal pulse. The coincidence we are interested in is the right peak at the plus time-delay. We integrate over a range of 0.8 ns around the center of this peak (corresponding to the grey area shown in the figure) to obtain the total number of coincidences.

To characterize the background coincidence, we block the gate pulse and perform the same two-photon coincidence measurements. The red line shows the measured results. The left peak disappears as expected, but the right peak remains noticeable. This is because a small fraction of the strong optical rotation pulses leaks into the gate detection path despite of the spectral filter. Since each rotation pulse has very short time delay with respect to the gate pulse (~150 ps), the coincidence detection of a rotation photon by the gate path and a signal photon by the signal path lies in the same delay time range (the grey area). We subtract this background to obtain an accurate measurement of signal intensity conditioned on detecting a gate photon.

We process the data for Fig. 2C of the main text using the same method. The only difference is that the signal pulse is longer (1.34 ns) for the data in Fig. 2C, and the delay between the gate and signal pulses becomes 4 ns. We integrate the number of coincidences measured over a range of 1.4 ns around the center of the right peak and subtract the background measured in the same way as described above.

3. Device characterization

3.1 Verification of a negatively charged quantum dot

To verify that the quantum dot contains a spin, we measure the photoluminescence of the device as a function of the applied magnetic field. Figure S5A shows the photoluminescence spectra of the device when we excite the sample with a 780-nm continuous-wave laser. At 0 T, the quantum dot is almost resonant with the cavity. Due to strong coupling, the emission spectrum shows a quantum-dot-like polariton (labeled as QD) and a cavity-like polariton (labeled as Cavity). As we increase the magnetic field, the quantum-dot-like polariton splits into four peaks, corresponding to the transitions $\sigma_1$ to $\sigma_4$ as shown in Fig. 1A of the main text, demonstrating that this dot is charged. Note that as we increase the magnetic field, transition $\sigma_1$ initially red shifts towards the cavity, and then blue shift due to the diamagnetic effect. For all the measurements reported in the manuscript, we apply a magnetic field of 5.5 T, where transition $\sigma_1$ is resonant with the cavity mode.

To determine whether the dot is positively or negatively charged, we measure the Lande g-factor of the ground states of the quantum dot. From the Ramsey fringes shown in Fig. 2 of the main text, we extract that the spin precession period to be $T = 25.5$ ps at a magnetic field of 5.5 T. Thus, the energy splitting between the two ground states are given by $\Delta_e/2\pi = 1/T = 39.2$ GHz. From this measurement, we calculate the Lande g-factor to be $g_l = 0.51$ using the relation $g_l = \hbar\Delta_e/\mu_B B$. This value is consistent with the typically reported numbers for a quantum dot containing a single electron that range from 0.4 to 0.6 (*19, 33-39*). Positively charged quantum dots containing a hole spin exhibit a Lande g-factor below 0.3 (*40, 41*), much smaller than our measured values, indicating that our spin is originating from an additional electron in the quantum dot.

3.2 Measurement of device parameters

We extract the cavity quantum electrodynamics parameter of the device from the cavity reflection spectrum. To obtain the cavity energy decay rate $\kappa$, we detune the cavity away from the quantum dot using nitrogen gas deposition and measure the bare cavity spectrum. We excite the cavity with a tunable continuous-wave laser in the right-circular polarization basis, and measure the reflected intensity in the left-circular polarization basis. The blue circles in Fig. S5B show the measured reflected laser intensity as we sweep the frequency of the laser across the cavity mode. We numerically fit the measured data to a function given by

$$S_{out}(\omega) = A \cdot \left| \frac{\kappa/2}{\kappa/2 + i(\omega - \omega_c)} \right|^2 + B, \tag{1}$$

where $S_{out}(\omega)$ is the intensity of the reflected laser at frequency $\omega$, $\omega_c$ is the frequency of the cavity mode, $\kappa$ is the cavity energy decay rate, $A$ is an overall intensity scaling factor that depends on the incident laser intensity and collection efficiency, and $B$ is the dark counts of the detector. From the numerical fit (blue solid line), we determine the cavity energy decay rate to be $\kappa/2\pi = 33.5 \pm 0.6 \text{ GHz}$.

Another important parameter about the device is the interference contrast $\alpha$, defined as $\alpha = \kappa_{ex}/\kappa$, where $\kappa_{ex}$ is the cavity energy decay rate to the reflected mode. For an ideal single-sided cavity $\alpha = 1$, but realistic cavities may suffer from intra-cavity losses which serve to degrade the interference. We could directly obtain $\alpha$ from the empty cavity reflection spectrum measured at the co-polarization basis. The reflected intensity in this case is given by

$$S_{out}(\omega) = A' \cdot \left| 1 - \alpha \frac{\kappa/2}{\kappa/2 + i(\omega - \omega_c)} \right|^2 + B, \tag{2}$$

where $A'$ relates with $A$ by the equation $A = A' \cdot \alpha^2$. On resonance, the expression takes on the simplified form $S_{out}(\omega_c) = A' \cdot |1 - \alpha|^2 + B$. Thus, we can directly infer $\alpha$ from the degree of suppression at the cavity resonance. The red diamonds in Fig. S5B show the measured bare cavity spectrum at such a polarization basis. We obtain that $\alpha = 0.92 \pm 0.01$ by fitting the measured data to the model described by Eq. (2) (red solid line).

To obtain the coupling strength $g$ between quantum dot transition $\sigma_1$ and the cavity, we tune the cavity back to its original frequency, and apply a magnetic field of 5.5 T such that transition $\sigma_1$ is resonantly coupled with the cavity. We excite the cavity with a tunable continuous-wave laser in the right-circular polarization basis, and measure its reflected intensity in the left-circular polarization basis. We also use another continuous-wave laser to resonantly excite transition $\sigma_4$ to initialize the spin in the spin-down state through optical pumping. The blue circles in Fig. S5C shows the measured spectrum. We numerically fit the measured spectrum to a master equation that accounts for dissipation and dephasing as described in our previous paper (*17*) (blue solid line), from which we extract the coupling strength between transition $\sigma_1$ and the cavity to be $g_1/2\pi = 10.7 \pm 0.2 \text{ GHz}$, and the dipole decoherence rate of transition $\sigma_1$ to be $\gamma_1/2\pi = 3.5 \pm 0.3 \text{ GHz}$. Based on these values, we calculate the device cooperativity to be $C = 2g^2/\kappa\gamma = 1.96 \pm 0.19$. We also obtain the coupling strength between transition $\sigma_2$ and the cavity to be $g_2/2\pi = 6.2 \pm 0.7 \text{ GHz}$, and the dipole decoherence rate of transition $\sigma_2$ to be $\gamma_2/2\pi = 6.9 \pm 1.7 \text{ GHz}$.

**Supplementary Text**

1. <u>Analytical description of the single-photon switch and transistor operations</u>

As described in the main text, we start our operation by preparing the quantum dot in a superposition of its spin ground states given by $(|\uparrow\rangle+|\downarrow\rangle)/\sqrt{2}$ by using an initialization pulse followed by a $\pi/2$ rotation. The system then freely evolves for a time $\tau$. If we set $\tau$ to be an integer number plus one half of spin precession period, the spin will evolve to the state $(|\uparrow\rangle-|\downarrow\rangle)/\sqrt{2}$. Then we inject a gate pulse which is quasi-resonant with the cavity mode and polarized parallel with the cavity. If the gate pulse contains a single photon, the spin-photon wavefunction transforms as $|1\rangle \otimes (|\uparrow\rangle-|\downarrow\rangle) \to r_\uparrow |1\rangle|\uparrow\rangle - r_\downarrow |1\rangle|\downarrow\rangle$, where $|1\rangle$ denotes the single-photon state, and $r_\uparrow$ and $r_\downarrow$ are the cavity reflection coefficients for the spin-up and spin-down states respectively, given by $r_\uparrow = -(2\alpha-1)$ and $r_\downarrow = 1 - 2\alpha/(1+C)$ respectively (*42*). For an ideal device where $C \gg 1$ and $\alpha = 1$, we have $r_\uparrow = -1$ and $r_\downarrow = 1$. Thus, the spin-photon wavefunction after the reflection of the gate photon is given by $|\psi_{out}\rangle = |1\rangle \otimes (|\downarrow\rangle+|\uparrow\rangle)$. Conditioned on detecting a reflected gate photon, the spin flips from state $(|\uparrow\rangle-|\downarrow\rangle)/\sqrt{2}$ to state $(|\uparrow\rangle+|\downarrow\rangle)/\sqrt{2}$. This operation can be viewed as a conditional spin flip along the *x*-axis. A subsequent $\pi/2$ pulse rotates the spin, transforming the operation to a conditional spin flip along the original spin basis. Thus, the spin occupies the spin-down state in the absence of the gate photon, and spin-up state with the gate photon. We prepare the signal field in the right-circular polarization basis, and detect its reflected intensity in the left-circular polarization basis. The transmittance through the detection polarizer is given by $T_{\uparrow(\downarrow)} = |1 - r_{\uparrow(\downarrow)}|^2 / 4$ (*43*). For an ideal device, we thus have $T_\uparrow = 1$ and $T_\downarrow = 0$. Therefore, a single gate photon could change the transmission of the signal field from 0 to 1. Similarly, if we set the free evolution time $\tau$ to be an integer number of spin precession period, a single gate photon could change the transmission of the signal field from 1 to 0. Finite cooperativity and non-unity α will degrade the switching contrast, but will not affect the fundamental working principle of the device.

2. <u>Measurement of signal photon number</u>

In this section, we show how we measure the average number of signal photons per pulse contained within the transverse spatial mode of the cavity. Since we can easily measure the average power of the signal pulse before the objective lens, all we need to know is the coupling efficiency from the excitation fiber to the transverse spatial mode of the cavity. To do this, we apply a weak coherent field with the same frequency as the signal field and from the same excitation fiber, and utilize the backaction of this field on the spin to determine the number of photons in this pulse coupled to the cavity and thus the coupling efficiency. For a pure spin state initialized in the spin-down state, if we apply two $\pi/2$ rotations with a variable delay time τ, the spin-up population will oscillate as a function of τ with a visibility of 1. However, if we apply a weak coherent field in between the two $\pi/2$ rotations, the spin will become entangled with the polarization of the photons that couple to the cavity spatial mode. Thus, the subsystem of the spin is no longer a pure state, which will result in a reduced visibility in the Ramsey interference fringes. From the level of degradation in the visibility, we could deduce the photon number in the applied coherent field that couple to the cavity, and thus the coupling efficiency.

Figure S6A shows the pulse sequence for our measurements. We first prepare the spin in a superposition state $(|\uparrow\rangle+|\downarrow\rangle)/\sqrt{2}$ by applying an initialization pulse followed by a $\pi/2$ rotation. We then send in a coherent pulse that is prepared in the right-circular polarization. After that, we send a second $\pi/2$ rotation pulse that rotates the spin back to the up-down basis. We could statistically read out the spin-up population from the intensity of the transition $\sigma_2$ emission induced by the optical pumping pulse in the next cycle (*19*).

Figure S6B shows the emission intensity from transition $\sigma_2$ as we vary the delay time $\tau$ between the two $\pi/2$ rotation pulses. The blue circles show the case when we block the middle coherent pulse, and the red diamonds show the case when we have the middle coherent pulse, which has an average power of 217.5 pW measured before the objective lens. In both cases, we observe oscillations in the emission intensity due to Ramsey interferences (*19*). However, the visibility of the Ramsey fringe degrades by 22% when we have the middle pulse. Here we define the visibility as $V=(I_{max}-I_{min})/(I_{max}+I_{min})$, where $I_{max}$ and $I_{min}$ are the maximum and minimum emission intensities respectively. The black squares show the case when we increase the average power of the middle coherent pulse to 380.6 pW. The visibility further degrades as we increase the power of the applied pulse.

The level of reduction in the visibility allows us to calculate the average number of photons per pulse coupled to the transverse spatial mode of the cavity. To accurately model the Ramsey visibility for a realistic device, we solve the system master equation. The system we consider is a four-level quantum dot coupled to a single-mode cavity, and the cavity is driven by a coherent pulse. We set the initial state of the quantum dot to be a superposition between the two spin ground states. The visibility in the Ramsey interference measurement corresponds to the length of the spin Bloch vector after it interacts with the applied coherent field, given by $V=\sqrt{2Tr(\rho_f^2)-1}$, where $V$ is the visibility, and $\rho_f$ is the density matrix of the system after the coherent pulse dies out.

We solve the system master equation given by $d\rho/dt=-(i/\hbar)[\hat{\mathbf{H}},\rho]+\hat{\mathbf{L}}\rho$, where $\rho$ is the density matrix of the system, $\hat{\mathbf{H}}$ is the system Hamiltonian, and $\hat{\mathbf{L}}$ is the Liouvillian superoperator that accounts for non-unitary evolution due to all dissipative mechanisms. We define the center frequency of the applied coherent pulse as $\omega$, and express the Hamiltonian in a reference with respect to this frequency. We express $\hat{\mathbf{H}}$ as $\hat{\mathbf{H}} = \hat{\mathbf{H}}_0 + \hat{\mathbf{H}}_{int} + \hat{\mathbf{H}}_d$, where

$$\hat{\mathbf{H}}_0 = \hbar(\omega_c - \omega)\hat{\mathbf{a}}^\dagger\hat{\mathbf{a}} + \hbar(\omega_x - \omega)\hat{\boldsymbol{\sigma}}_1^\dagger\hat{\boldsymbol{\sigma}}_1 + \hbar(\omega_x - \omega + \Delta_h)\hat{\boldsymbol{\sigma}}_2^\dagger\hat{\boldsymbol{\sigma}}_2 - \hbar\Delta_e\hat{\boldsymbol{\sigma}}_2\hat{\boldsymbol{\sigma}}_2^\dagger, \tag{3}$$

$$\hat{\mathbf{H}}_{int} = \hbar g_1\left(\hat{\mathbf{a}}\hat{\boldsymbol{\sigma}}_1^\dagger + \hat{\boldsymbol{\sigma}}_1\hat{\mathbf{a}}^\dagger\right) + \hbar g_1\left(\hat{\mathbf{a}}\hat{\boldsymbol{\sigma}}_4^\dagger + \hat{\boldsymbol{\sigma}}_4\hat{\mathbf{a}}^\dagger\right) + i\hbar g_2\left(\hat{\mathbf{a}}\hat{\boldsymbol{\sigma}}_2^\dagger - \hat{\boldsymbol{\sigma}}_2\hat{\mathbf{a}}^\dagger\right) + i\hbar g_2\left(\hat{\mathbf{a}}\hat{\boldsymbol{\sigma}}_3^\dagger - \hat{\boldsymbol{\sigma}}_3\hat{\mathbf{a}}^\dagger\right), \tag{4}$$

$$\hat{\mathbf{H}}_d = i\hbar\sqrt{\frac{\kappa_{ex}}{2}} \cdot \sqrt{n_{in}G(t)} \cdot \left(\hat{\mathbf{a}}^\dagger - \hat{\mathbf{a}}\right). \tag{5}$$

In Eqs. (3) – (5), $\hat{\boldsymbol{\sigma}}_1$, $\hat{\boldsymbol{\sigma}}_2$, $\hat{\boldsymbol{\sigma}}_3$ and $\hat{\boldsymbol{\sigma}}_4$ are the lowering operators for the quantum dot transitions $\sigma_1$, $\sigma_2$, $\sigma_3$ and $\sigma_4$ respectively, and $\hat{\mathbf{a}}$ is the photon annihilation operator for the cavity mode. The remaining parameters are defined as follows: $\omega_c$ is the cavity resonant frequency, $\omega_x$ is the frequency of transition $\sigma_1$, $\Delta_e$ is the Zeeman splitting between the two electron ground states, $\Delta_h$ is the Zeeman splitting between the two excited trion states, $g_1$ is the coupling strength between the cavity and transitions $\sigma_1$ and $\sigma_4$ (they share the same coupling strength due to the same polarization selection rule), $g_2$ is the coupling strength between the cavity and transitions $\sigma_2$ and $\sigma_3$, $\kappa_{ex}$ is the cavity energy decay rate to the reflected mode, $n_{in} = \dfrac{\eta P}{\hbar \omega R}$ is the average number of photons per pulse in the applied coherent pulse that couple to the cavity, $P$ is the average power of the applied coherent pulse measured before the objective lens, $R = 76$ MHz is the repetition rate of the applied coherent pulse, $\eta$ is the coupling efficiency from the incident fiber mode to the transverse spatial mode of the cavity, and $G(t)$ is the time varying intensity of the applied coherent pulse, satisfying $\int_{-\infty}^{\infty} G(t)dt = 1$. Note that the coupling between the cavity and transition $\sigma_2 (\sigma_3)$ has a phase shift of $\pi/2$ compared with the coupling between the cavity and transition $\sigma_1 (\sigma_4)$ due to the selection rules (*44*).

The Liouvillian superoperator $\hat{L}$ accounts for all nonunitary Markovian processes including spontaneous emission, dephasing of the excited trion states and decay of the cavity field. This operator is given by

$$\hat{L} = \kappa D(\hat{\mathbf{a}}) + \gamma_1 D(\hat{\boldsymbol{\sigma}}_1) + \gamma_2 D(\hat{\boldsymbol{\sigma}}_2) + \gamma_3 D(\hat{\boldsymbol{\sigma}}_3) + \gamma_4 D(\hat{\boldsymbol{\sigma}}_4) + 2\gamma_{d1} D(\hat{\boldsymbol{\sigma}}_1^\dagger\hat{\boldsymbol{\sigma}}_1) + 2\gamma_{d2} D(\hat{\boldsymbol{\sigma}}_2^\dagger\hat{\boldsymbol{\sigma}}_2), \tag{6}$$

where $D(\hat{O})\rho = \hat{O}\rho\hat{O}^\dagger - 1/2\hat{O}^\dagger\hat{O}\rho - 1/2\rho\hat{O}^\dagger\hat{O}$ is the general Lindblad superoperator for the collapse operator $\hat{O}$. In Eq. (6), $\gamma_1$, $\gamma_2$, $\gamma_3$ and $\gamma_4$ are the quantum dot spontaneous emission rate of transitions $\sigma_1$, $\sigma_2$, $\sigma_3$ and $\sigma_4$ respectively, and $\gamma_{d1}$ and $\gamma_{d2}$ are the pure dephasing rates for the two trion states respectively.

We set the initial state of the system as $\rho_0 = \rho_{qd} \otimes \rho_{ph}$, where $\rho_{qd} = (|\uparrow\rangle + |\downarrow\rangle)(\langle\uparrow| + \langle\downarrow|)/2$ is the density matrix for the quantum dot initial state, and $\rho_{ph} = |vac\rangle\langle vac|$ is the density matrix for the cavity photons where $|vac\rangle$ represents the vacuum state. We set all the parameters as the experimentally measured values, given by $\omega_c = \omega_x = \omega$, $\Delta_e/2\pi = 39.2\,\text{GHz}$, $\Delta_h/2\pi = 19.0\,\text{GHz}$, $g_1/2\pi = 10.7\,\text{GHz}$, $g_2/2\pi = 6.2\,\text{GHz}$, $\kappa/2\pi = 33.5\,\text{GHz}$, $\alpha = \kappa_{ex}/\kappa = 0.92$, $\gamma_{d1}/2\pi = 3.5\,\text{GHz}$, $\gamma_{d2}/2\pi = 6.9\,\text{GHz}$, and $\gamma_1/2\pi = \gamma_2/2\pi = \gamma_3/2\pi = \gamma_4/2\pi = 0.1\,\text{GHz}$. The applied coherent pulse can be well modeled by a Gaussian function, thus we set $G(t)$ as

$$G(t) = \frac{2\sqrt{2\ln 2}}{\sqrt{2\pi} t_{fwhm}} \exp\left(-4\ln 2 \cdot \left(\frac{t - t_0}{t_{fwhm}}\right)^2\right),$$

where $t_{fwhm} = 63\,\text{ps}$ is the full width at half maximum of the gate pulse, and $t_0$ is the peak time of the pulse, which we set to be $t_0 = 20 t_{fwhm}$ so that at $t = 0$ the amplitude of the applied coherent pulse is nearly zero. We numerically solve the system density matrix, and obtain the final density matrix of the system $\rho_f$ at $t - t_0 \gg t_{fwhm}$.

We note that in our measurement, even in the absence of the applied coherent pulse, the visibility of the spin Ramsey fringe is not 1. This is due to imperfect spin initialization and rotations, and spin decoherence during the two rotation pulses. We use spin fidelity $F$ to characterize these imperfections, defined as $F = \langle \phi(\tau)|\rho_s(\tau)|\phi(\tau)\rangle$, where $\phi(\tau)$ is the ideal spin state after the two rotation pulses with a delay time $\tau$, given by $\phi(\tau) = \cos(\Delta_e \tau/2)|\uparrow\rangle - \sin(\Delta_e \tau/2)|\downarrow\rangle$, and $\rho_s(\tau)$ is the actual density matrix of the spin after the two rotation pulses without the applied coherent field. In general, the fidelity depends on the delay time $\tau$. But in our experiment, we vary $\tau$ within less than 100 ps, which is much shorter than the spin coherence time of ~ 2 ns. Thus, we could treat $F$ as one number that is independent of $\tau$. We calculate the visibility using $\bar{V} = (2F - 1)V$, where $\bar{V}$ is the Ramsey interference visibility when taking into account imperfect spin initialization and rotations, and spin decoherence during the two rotation pulses.

The black squares in Fig. S6C show the measured visibility in the spin Ramsey interference as we increase the power of the applied coherent field. The solid line shows the numerically calculated visibility $\bar{V}$, with $F$ and $\eta$ being the only free fitting parameters. From the numerical fit (solid line), we determine that $F = 0.747 \pm 0.006$, and $\eta = (3.16 \pm 0.35)\%$.

3. Numerical calculation of transmittance of the signal field

In this section, we show how we extract the transmittance of the signal field from the measured intensity. The transmittance and the measured intensity obeys the simple relation given by $I(\tau) = A \cdot T(\tau)$, where $\tau$ is the delay time between the two spin rotation pulses, $I(\tau)$ is the measured signal field intensity when the two rotation pulses are separated by a delay time $\tau$, $T(\tau)$ is the transmittance of the signal field when the two rotation pulses are separated by a delay time $\tau$, and $A$ is a scaling factor that depends on the input signal field intensity and the collection and detection efficiency. Since in general the scaling factor $A$ is difficult to obtain, we cannot directly calculate the transmittance from one measured intensity. However, we could obtain $A$ and $T(\tau)$ from all the measured intensities when we vary the delay time $\tau$.

We first focus on the measurement without the gate pulse. The transmittance $T(\tau)$ is given by

$$T(\tau) = P_\uparrow(\tau) T_\uparrow + P_\downarrow(\tau) T_\downarrow, \tag{7}$$

where $T_\uparrow$ and $T_\downarrow$ are the transmittance of the signal field when the spin is in spin-up and spin-down states respectively, and $P_\uparrow(\tau)$ and $P_\downarrow(\tau)$ are the probabilities of the spin being in spin-up and spin-down states after the two rotation pulses that are separated by a delay time $\tau$. The values of $P_\uparrow(\tau)$ and $P_\downarrow(\tau)$ in the absence of the gate field are given by

$$P_\uparrow(\tau) = F \cos^2\left(\frac{\Delta_e \tau}{2}\right) + (1-F) \sin^2\left(\frac{\Delta_e \tau}{2}\right), \tag{8}$$

$$P_\downarrow(\tau) = (1-F) \cos^2\left(\frac{\Delta_e \tau}{2}\right) + F \sin^2\left(\frac{\Delta_e \tau}{2}\right), \tag{9}$$

where $F$ is the fidelity of the spin after the two rotation pulses as defined in Section 2.

To calculate $T_\uparrow$ and $T_\downarrow$, we solve the system master equation given by $d\rho_{\uparrow(\downarrow)}/dt = -i/\hbar \left[\hat{\mathbf{H}}, \rho_{\uparrow(\downarrow)}\right] + \hat{\mathbf{L}} \rho_{\uparrow(\downarrow)}$, where $\rho_{\uparrow(\downarrow)}$ is the density matrix of the system when the quantum dot is initially at spin-up (spin-down) state, $\hat{\mathbf{H}}$ is the system Hamiltonian given by Eq. (3) – (5), and $\hat{\mathbf{L}}$ is the Liouvillian superoperator that is given by Eq. (6). The average number of photons in the reflected signal field is given by

$$m_{\uparrow(\downarrow)} = \frac{\kappa_{ex}}{2} \int_{-\infty}^{\infty} Tr\left(\rho_{\uparrow(\downarrow)}(t) \hat{\mathbf{a}}^\dagger \hat{\mathbf{a}}\right) dt, \tag{10}$$

where $m_{\uparrow(\downarrow)}$ is the number of reflected photons when the spin is initially at spin-up (spin-down) state. The transmittance $T_\uparrow$ and $T_\downarrow$ are given by $T_{\uparrow(\downarrow)} = m_{\uparrow(\downarrow)}/n_{in}$, where $n_{in} = N_s$ is the average number of photons per pulse in the signal field contained within the transverse spatial mode of the cavity. We set all the parameters as the experimentally measured values, given by $\omega_c = \omega_x = \omega$, $\Delta_e/2\pi = 39.2\,\text{GHz}$, $\Delta_h/2\pi = 19.0\,\text{GHz}$, $g_1/2\pi = 10.7\,\text{GHz}$, $g_2/2\pi = 6.2\,\text{GHz}$, $\kappa/2\pi = 33.5\,\text{GHz}$, $\alpha = \kappa_{ex}/\kappa = 0.92$, $\gamma_{d1}/2\pi = 3.5\,\text{GHz}$, $\gamma_{d2}/2\pi = 6.9\,\text{GHz}$, and $\gamma_1/2\pi = \gamma_2/2\pi = \gamma_3/2\pi = \gamma_4/2\pi = 0.1\,\text{GHz}$. Our signal pulse can be well modeled by a Gaussian function, thus we set $G(t)$ as $G(t) = \frac{2\sqrt{2\ln 2}}{\sqrt{2\pi} t_{fwhm}} \exp\left(-4\ln 2 \cdot \left(\frac{t-t_0}{t_{fwhm}}\right)^2\right)$, where $t_{fwhm}$ is the full width at half maximum of the gate pulse, and $t_0$ is the peak time of the pulse. For the data shown in Fig. 2A and 2B, we have $t_{fwhm} = 63$ ps. For the data shown in Fig. 2C, we have $t_{fwhm} = 1.34$ ns. We always set $t_0$ to be $t_0 = 20 t_{fwhm}$ so that at $t = 0$ the signal pulse amplitude is nearly zero. For Fig. 2A where $N_s = 0.42$, we calculate $T_\uparrow$ and $T_\downarrow$ to be 0.79 and 0.37 respectively. For Fig. 2C where $N_s = 4.4$, 10.9 and 23.0, we calculate $T_\uparrow$ to be 0.81, 0.80 and 0.78 respectively, and $T_\downarrow$ to be 0.37, 0.47 and 0.57 respectively.

We numerically fit the measured intensities to the model given by $I(\tau) = A \cdot T(\tau)$, with the scaling factor $A$ and the spin fidelity $F$ being the only free fitting parameters. From the numerical fit to Fig. 2A, we obtain that $F = 0.783 \pm 0.009$, which matches well with the value obtained in Section 2 ($F = 0.747 \pm 0.006$) by measuring the visibility of the Ramsey fringes from the quantum dot fluorescence. The spin fidelities extracted from the red diamonds (in the absence of the gate pulse) in Fig. 2C have nearly identical values with Fig. 2A, given by $0.759 \pm 0.004$, $0.739 \pm 0.007$, and $0.787 \pm 0.010$ from the upper to the lower panel. We could thus calculate the transmittance of the signal field $T(\tau)$ using Eq. (7) - (9).

We follow the same formalism to calculate the transmittance of the signal field conditioned on detecting a gate photon. The measured conditional intensity $I'(\tau)$ is given by $I'(\tau) = A' \cdot T'(\tau)$, where $A'$ is another scaling factor, and $T'(\tau)$ is the transmittance of the signal field conditioned on detecting a gate photon, given by

$$T'(\tau) = P'_\uparrow(\tau) T_\uparrow + P'_\downarrow(\tau) T_\downarrow, \tag{11}$$

where $P'_\uparrow(\tau)$ and $P'_\downarrow(\tau)$ are the probabilities of the spin being in spin-up and spin-down states conditioned on detecting a gate photon after the two rotation pulses that are separated by a delay time $\tau$. The values of $P'_\uparrow(\tau)$ and $P'_\downarrow(\tau)$ are given by

$$P'_\uparrow(\tau) = F' \sin^2\left(\frac{\Delta_e \tau}{2}\right) + (1-F') \cos^2\left(\frac{\Delta_e \tau}{2}\right), \tag{12}$$

$$P'_\downarrow(\tau) = (1-F') \sin^2\left(\frac{\Delta_e \tau}{2}\right) + F' \cos^2\left(\frac{\Delta_e \tau}{2}\right). \tag{13}$$

In Eq. (12) and (13), $F'$ is the spin fidelity conditioned on detecting a gate photon, which accounts for both spin decoherence due to imperfect coherent spin manipulations, and imperfect spin-photon interactions (e.g. a single gate photon does not create an exact $\pi$ phase shift between spin-up and spin-down, or the spin decoheres due to the small probability of interacting with multiple photons in the gate pulse).

We again numerically fit the measured conditional intensities to the model given by $I'(\tau) = A' \cdot T'(\tau)$, with the scaling factor $A'$ and the spin fidelity $F'$ being the only free fitting parameters. In the numerical fit to the conditional intensities obtained in Fig. 2B, we obtain that $F' = 0.709 \pm 0.031$. This value degrades slightly from the spin fidelity in the absence of the gate pulse given by $F = 0.783 \pm 0.009$, due to the imperfect interactions between the spin and the gate photon. We calculate the transmittance of the signal field $T'(\tau)$ using Eq. (11)-(13).

4. Calculation of switching contrast

We define switching contrast $\xi$ of the single-photon switch as the change in the transmittance of the signal field induced by a single gate photon. As discussed in the main text, the switching contrast achieves the same maximum value at delay condition $a$ and $b$. We therefore calculate the switching contrast at delay condition $a$ or $b$, given by $\xi = T_g^{(a)} - T_{ng}^{(a)}$ or $\xi = T_{ng}^{(b)} - T_g^{(b)}$, where $T_{ng}^{(a)}$ and $T_{ng}^{(b)}$ are the transmittance of the signal field in the absence of the gate photon when the delay time between the two rotation pulses are at condition $a$ and $b$ respectively, and $T_g^{(a)}$ and $T_g^{(b)}$ are the transmittance of the signal field with a single gate photon when the delay time between the two rotation pulses are at condition $a$ and $b$ respectively.

We directly obtain $T_{ng}^{(a)}$ or $T_{ng}^{(b)}$ from the measured values shown in Fig. 2 of the main text. For $T_g^{(a)}$ and $T_g^{(b)}$, since we use a weak coherent field as the gate, the measured values are degraded from the actual transmittance with a single gate photon due to the possibility of the multi-photon events in the gate pulse. In our experiment, we use a gate field with an average power of 217.5 pW before the objective lens. As shown in Section 2 of the supplementary text, such a coherent field will shrink the length of the spin Bloch vector from its original value with a factor $\beta = 0.78$. We thus correct the multi-photon events using $T_g^{(a,b)} = \left(\bar{T}_g^{(a,b)} - T_0\right)/\beta + T_0$, where $\bar{T}_g^{(a)}$ and $\bar{T}_g^{(b)}$ are the measured conditional transmittance when the delay time between the two rotation pulses are at condition $a$ and $b$ respectively, $T_0$ is the transmittance of the signal field when the spin is in a complete mixture state, which can be calculated as $T_0 = \left(\bar{T}_g^{(a)} + \bar{T}_g^{(b)}\right)/2$.

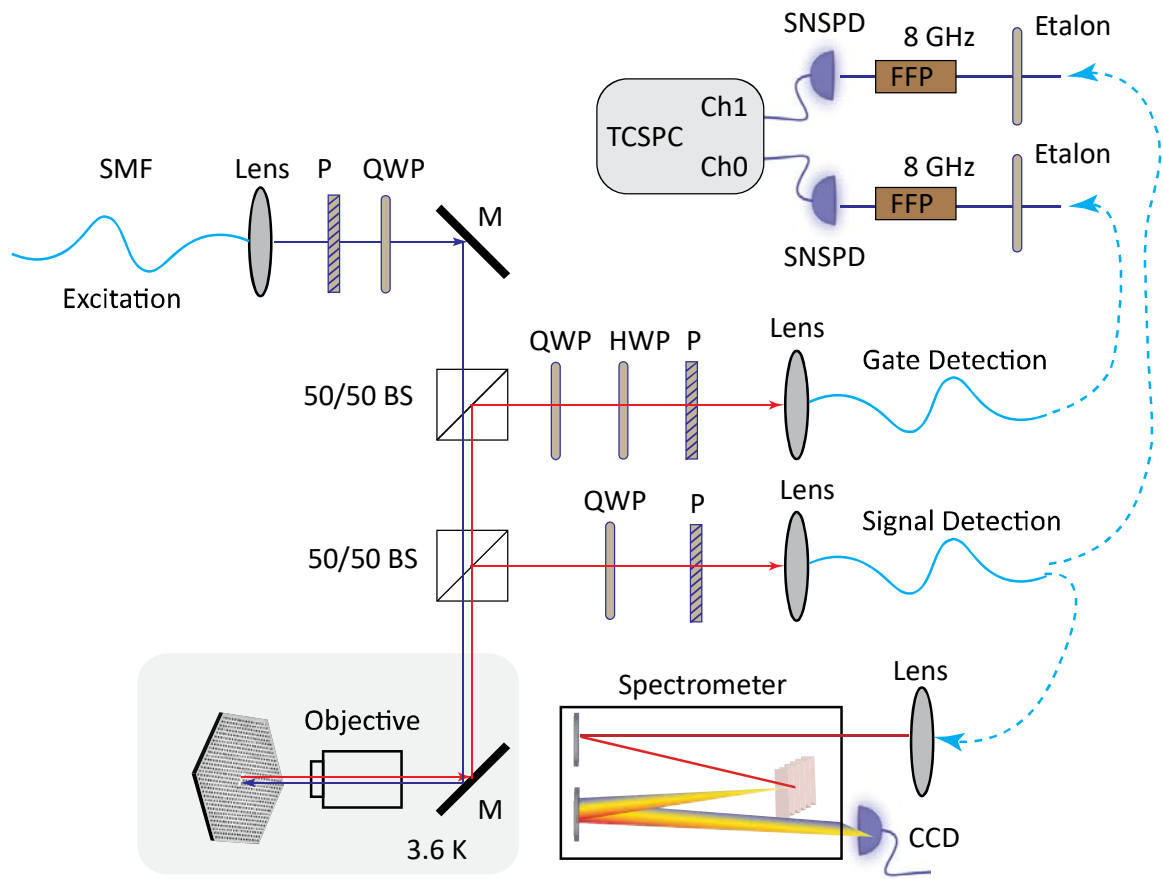

**Fig. S1.**
**Schematics of the whole experimental setup.** SMF, single mode fiber; P, polarizer; QWP, quarter wave plate; HWP, half wave plate; M, mirror; BS, beam splitter; TCSPC, time correlated single photon counting module; SNSPD, superconducting nanowire single photon detector; FFP, fiber Fabry-Perot filter; CCD, charged coupled device.

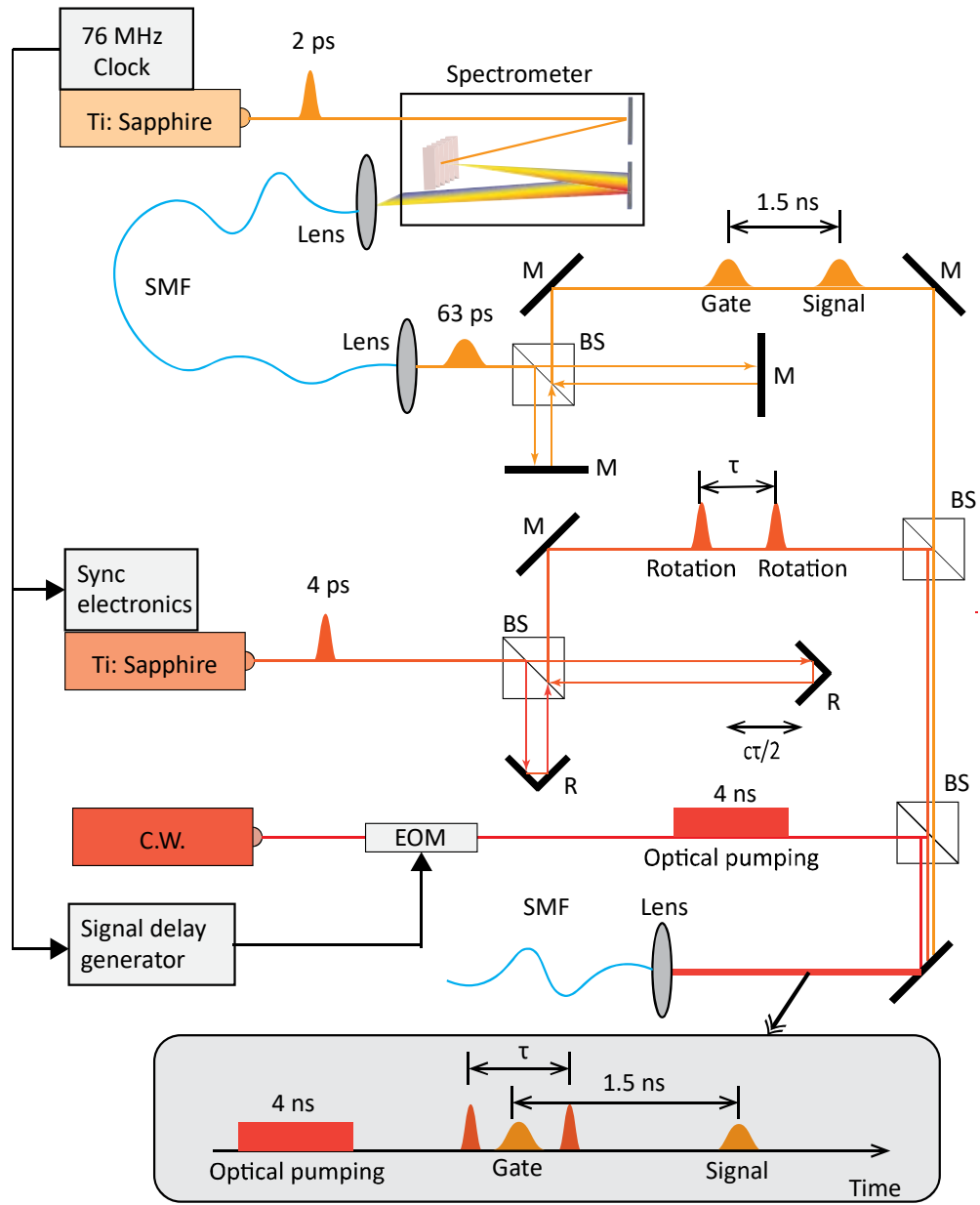

**Fig. S2.**

**Setup to generate the pulse sequence used in Fig. 2A and 2B of the main text.** SMF, single mode fiber; M, mirror; BS, beam splitter; C.W., continuous wave laser; EOM, electro-optic modulator; R, retro-reflector. The inset at the left bottom corner shows the schematic of the generated pulse sequence.

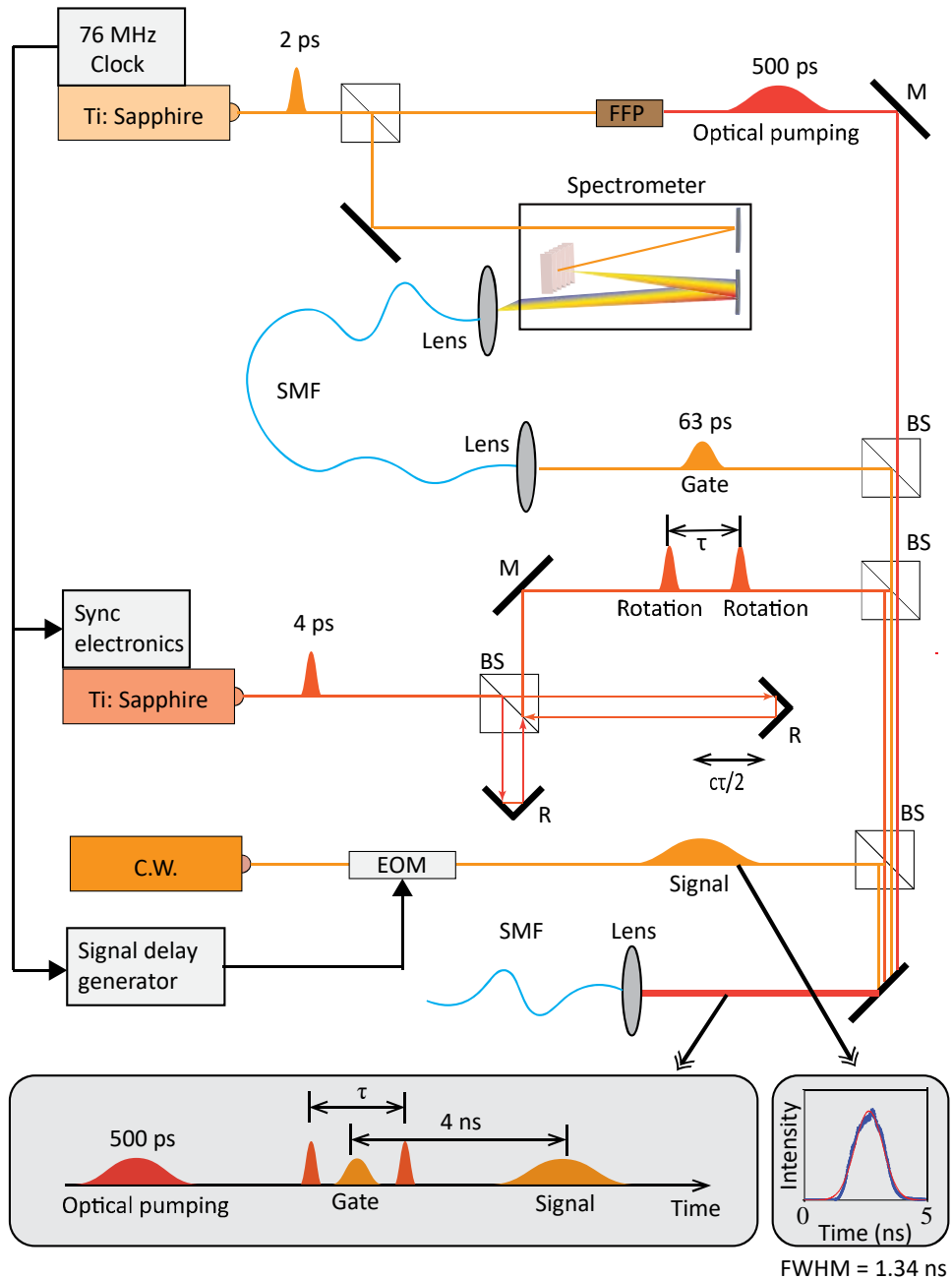

**Fig. S3.**

**Setup to generate the pulse sequence used in Fig. 2C of the main text.** SMF, single mode fiber; M, mirror; BS, beam splitter; C.W., continuous wave laser; EOM, electro-optic modulator; R, retro-reflector; FFP, fiber Fabry Perot filter. The inset at the left bottom corner shows the schematic of the generated pulse sequence. The inset at the right bottom corner shows the measured shape of the signal pulse. The blue dots show the measured data, and the red solid line shows a numerical fit to a Gaussian function with FWHM of 1.34 ns.

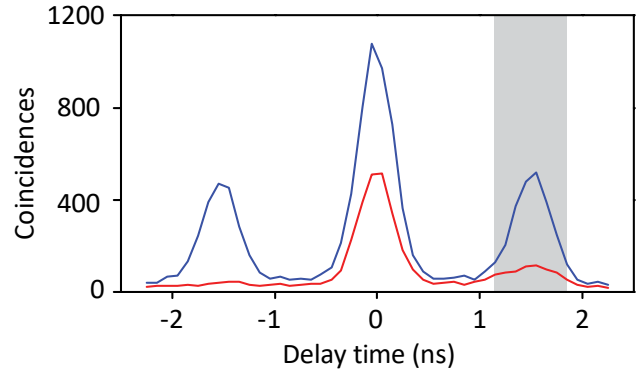

**Fig. S4.**

**Sample coincidence histogram we obtained to extract one of the data point shown in Fig. 2B of the main text.** The blue and red lines show the coincidence histogram in the presence and absence of the gate pulse respectively. The grey area shows the integration window we used to calculate the total number of coincidences between the gate and signal photon.

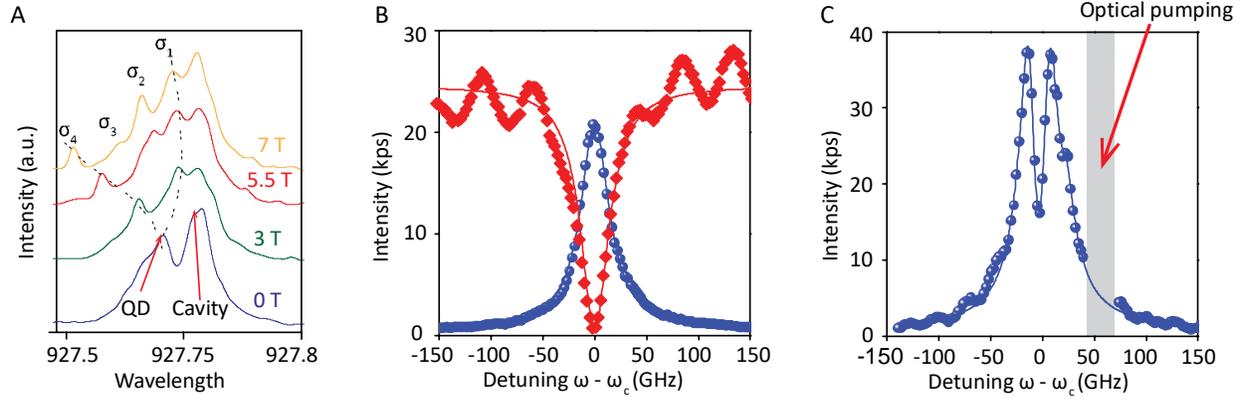

**Fig. S5.**

**Device characterization. A,** photoluminescence spectra of the device at several different magnetic fields. **B,** bare cavity reflection spectrum measured with cross-polarization (blue circles) and co-polarization (red diamonds) basis respectively. The blue and red solid line shows numerically calculated spectra. **C,** cross-polarized cavity reflection spectrum when transition of the quantum dot is resonant with the cavity. The blue circles show the measured data, and the blue solid line shows the numerically calculated spectrum.

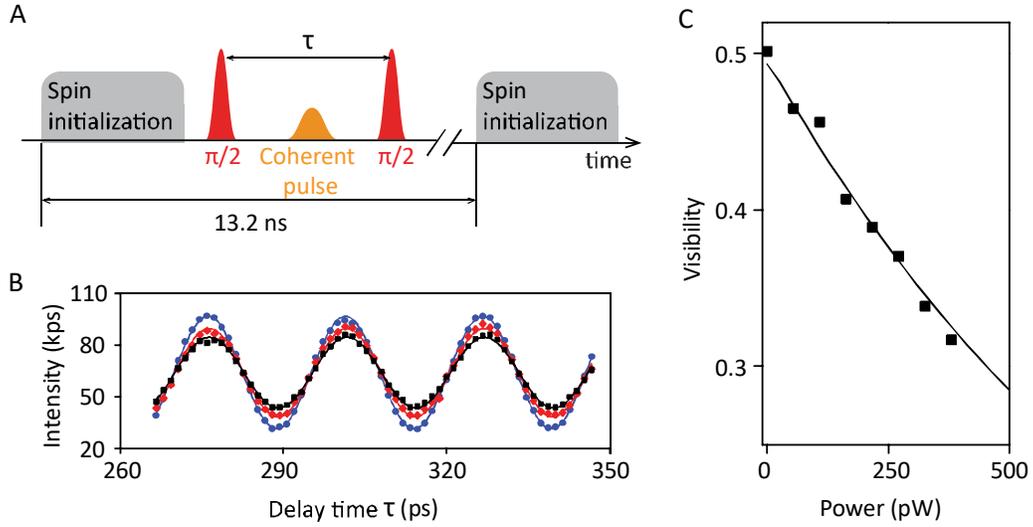

**Fig. S6.**

**Measurement of average photon number in a coherent pulse coupled to the transverse spatial mode of the cavity. A,** schematics of the measurement pulse sequence. **B,** intensity of the emission from transition in the absence of the coherent pulse (blue circles), in the presence of the coherent pulse with average power of 217.5 pW (red diamonds) and 380.6 pW (black squares) respectively, measured before the objective lens. The blue, red and black solid lines show numerically calculated values. **C,** extracted visibility of the Ramsey fringes as a function of the average power of the coherent pulse measured before the objective lens. Black squares show measured data, and black solid line shows a numerical fit to the solution of the system master equation.

18